\documentclass[]{raa}            
\usepackage{graphicx,times}
\usepackage{amssymb}
\usepackage{natbib}
\begin{document}
   \title{Asteroseismology of the DBV star CBS 114
}

\volnopage{ {\bf 0000} Vol.\ {\bf 0} No. {\bf XX}, 000--000}
\setcounter{page}{1}

\author{Y. H. Chen
\inst{1}}

\institute{\inst{1}  School of Physics and Electronical Science, Chuxiong Normal University, Chuxiong 675000, China; {yanhuichen1987@ynao.ac.cn\\}
~~~{\it }\\
\vs \no
{\small Received [0000] [July] [day]; accepted [0000] [month] [day] }}

\abstract{Asteroseismology is an unique and powerful tool to detect the internal structure of stars. CBS 114 is the sixth known pulsating DBV star. It was observed by Handler, Metcalfe, \& Wood at South African Astronomical Observatory over 3 weeks in 2001. Then, it was observed by Metcalfe et al. at Bohyunsan Optical Astronomy Observatory and McDonald Observatory respectively for 7 nights in 2004. Totally 2 triplets, 4 doublets, and 5 singlets were identified. The frequency splitting values are very different, from 5.2 $\mu$Hz to 11.9 $\mu$Hz, which may reflect differential rotations. We evolve grids of white dwarf models by \texttt{MESA}. Cores, added with He/C envelopes, of those white dwarf models are inserted into \texttt{WDEC} to evolve grids of DBV star models. With those DBV star models, we calculate eigenperiods. Those calculated periods are used to fit observed periods. A best-fitting model is selected. The parameters are $T_{eff}$ = 25000 K, $M_{*}$ = 0.740 $M_{\odot}$, and log($M_{He}/M_{*}$) = -4.5. With the massive stellar mass, the effective temperature is close to previous spectroscopic result. In addition, kinetic energy distributions are calculated for the best-fitting model. We find that the observed modes with large frequency splitting values are fitted by the calculated modes with much kinetic energy distributed in C/O core. After preliminary analysis, we suggest that the C/O core may rotate at least 2 times faster than the helium layer for CBS 114.
\keywords{stars: oscillations (including pulsations)-stars: individual (CBS 114)-white dwarfs} }

\authorrunning{Y. H. Chen}            
\titlerunning{Asteroseismology of the DBV star CBS 114}  
\maketitle

\section{Introduction}

White dwarfs are the final evolutionary stage of about 98\% of stars (Fontaine, Brassard, \& Bergeron 2001). Research on white dwarfs is of general significance. About 80\% of white dwarfs show hydrogen atmosphere (DA type) and about 20\% of them show helium atmosphere (DB type) (Bischoff-Kim \& Metcalfe 2011). Basically, there is no thermonuclear reaction for white dwarfs. They are cooling down by eradiating. Along the cooling track, there are DOV (around 75000 K to 170000 K), DBV (around 22000 K to 29000 K), and DAV (around 10800 K to 12270 K) instability strips (Winget \& Kepler 2008). The pulsation of white dwarfs makes it possible for us to study the internal structure of white dwarfs.

Observed eigenfrequencies carry internal structure information of a pulsating star. Asteroseismology can be used to detect the internal structure information. White dwarfs are $g$-mode pulsators. Radial order $k$, spherical harmonic degree $l$, and azimuthal number $m$ are used to characterize an eigenmode. Tassoul (1980) reported an asymptotic theory for $g$-modes as
\begin{equation}
\bar{\triangle \texttt{P}(l)}=\frac{2\pi^{2}}{\sqrt{l(l+1)}{\int_{0}}^{R}\frac{|N|}{r}dr}.
\end{equation}
\noindent In Eq.\,(1), $R$ is stellar radius. The Brunt-V\"ais\"al\"a frequency $N$ should be calculated in the absolute value. The parameter $k$ can not be identified from observations. It shows the nodes of standing wave inside a star. According to Eq.\,(1), relative radial orders may be counted from observations. Star rotation can cause frequency splitting phenomenon. An approximate formula between frequency splitting ($\delta\nu_{k,l}$) and rotational period ($P_{\rm rot}$) was reported by Brickhill (1975) as
\begin{equation}
m\delta\nu_{k,l}=\nu_{k,l,m}-\nu_{k,l,0}=\frac{m}{P_{\rm rot}}(1-\frac{1}{l(l+1)}).
\end{equation}
\noindent In Eq.\,(2), the parameter $m$ is taken to be integers from $-l$ to $l$. For $l$ = 1, modes with $m$ = -1, 0, +1 form a triplet. For $l$ = 2, modes with $m$ = -2, -1, 0, +1, +2 form a quintuplet. According to Eq.\,(2), the faster a star rotates, the larger the frequency splits. The frequency splitting can be used to study star rotation phenomenon. The different frequency splitting values can be used to study differential rotation phenomenon. Assuming low $k$ modes sense interior more than high $k$ modes, Winget et al. (1994) reported that the envelope of GD 358 rotated some 1.8 times faster than its core. Studying rotational splitting inversions of observational data for GD 358, Kawaler, Sekii, \& Gough (1999) reported that the core rotated faster than its envelope.

GD 358 is the prototype of DBV class. DBV stars are considered as a powerful probe of the energy loss rate for plasma neutrino (Winget et al. 2004). Winget \& Montgomery (2006) studied the energy loss rate for plasma neutrino by measuring period changing rate for hot DBV stars. In addition, research on DBV stars are helpful to check the star evolution theory. Nehner \& Kawaler (1995) tried to study evolutionary connections between PG 1159 stars and DBV stars. Diffusion has an important effect on the evolutions. The evolutionary scenario requires a Very Late Thermal Pulse (VLTP) burning off the residual hydrogen in the envelope during post-asymptotic-giant-branch (post-AGB) evolution. Iben et al. (1983) reported that about 20\% of post-AGB stars would experience a VLTP when they descend to the white dwarf cooling track. However, it is difficult to interpret the 'DB gap' phenomenon (Liebert et al. 1986). After the VLTP process, a few hydrogen (log($M_{H}$/$M_{*}$))$\sim$-11) left (Herwig et al. 1999). When the effective temperature cools to 30000 K, the helium convection dilutes the left hydrogen in the photosphere. The helium convection may interpret the 'DB gap' phenomenon. The asteroseismological study on DBV stars is significant and meaningful.

CBS 114 is the sixth known DBV star discovered by Winget \& Claver (1988, 1989). Handler, Metcalfe, \& Wood (2002) (HMW2002) reported their 65h of single-site time-resolved CCD photometry on CBS 114. Totally 7 independent modes were identified. Some of them might need corrections of daily alias 11.60 $\mu$Hz, as they reported. Taking C-core, O-core, and C/O-core into account, they did asteroseismological study on those 7 independent modes (assuming $l$ = 1). In order to obtain more modes to constrain fitting models for CBS 114, a dual-site campaign using 2\,m class telescopes in February 2004 was organized by Metcalfe et al. (2005). The previous 7 modes were recovered and 4 new ones were discovered. With pure carbon core and uniform He/C envelope, Metcalfe et al. (2005) made grids of white dwarf models to fit those 11 $l$ = 1 modes. The identified components show different frequency splitting values, from 5.2 $\mu$Hz to 11.9 $\mu$Hz, which may correspond to differential rotation phenomenon.

Paxton et al. (2011) reported Modules for Experiments in Stellar Astrophysics ($\texttt{MESA}$), which can evolve stars from pre-main sequence to white dwarf stage. We try to insert the core and envelope compositions of white dwarf models evolved by $\texttt{MESA}$ into an old program White Dwarf Evolution Code ($\texttt{WDEC}$) (Wood 1990). With this method, grids of DBV star models are evolved with element diffusion effect. We try to do asteroseismological study on CBS 114 and try to study its differential rotation. In Sect. 2, we introduce the input physics and model calculations. Model fittings on CBS 114 are showed in Sect. 3. In Sect. 3.1, 3.2, and 3.3, we study mode identifications, fitting results, and differential rotations for CBS 114 respectively. At last, we make some discussions and summarize our conclusions in Sect. 4.

\section{Input physics and model calculations}

\begin{table}
\begin{center}
\caption{Information about calculated models. In the header, $MS$, $WD(\texttt{MESA})$, and $WD(\texttt{WDEC})$ respectively denote stellar mass of main-sequences, white dwarfs evolved by $\texttt{MESA}$, and white dwarfs evolved by $\texttt{WDEC}$. $M_{core}(\texttt{MESA})$ is core mass of white dwarfs evolved by $\texttt{MESA}$. $M_{en}$ is He/C envelope mass. In the envelope, $X_{C(en)}$ and $X_{He(en)}$ denote the abundance of carbon and helium respectively.}
\begin{tabular}{lllll}
\hline
$MS$         &$WD(\texttt{MESA})$   &$M_{core}(\texttt{MESA})$ &$WD(\texttt{WDEC})$   &log($M_{en}/M_{core})(X_{C(en)},X_{He(en)}$) \\
\hline
($M_{\odot}$)&($M_{\odot}$)         &($M_{\odot}$)             &($M_{\odot}$)         &                                             \\
\hline
2.0          &0.580                 &0.558                     &0.550-0.575           &-1.714 (0.155,0.845)                         \\
2.8          &0.614                 &0.595                     &0.580-0.605           &-1.754 (0.185,0.815)                         \\
3.0          &0.633                 &0.616                     &0.610-0.635           &-1.938 (0.171,0.829)                         \\
3.2          &0.659                 &0.644                     &0.640-0.665           &-2.085 (0.145,0.855)                         \\
3.4          &0.689                 &0.675                     &0.670-0.695           &-1.844 (0.138,0.862)                         \\
3.6          &0.723                 &0.713                     &0.700-0.725           &-2.328 (0.046,0.954)                         \\
3.8          &0.751                 &0.742                     &0.730-0.755           &                                             \\
4.0          &0.782                 &0.774                     &0.760-0.785           &-2.430 (0.025,0.975)                         \\
4.5          &0.805                 &0.799                     &0.790-0.815           &-2.592 (0.055,0.945)                         \\
5.0          &0.832                 &0.827                     &0.820-0.850           &                                             \\
\hline
\end{tabular}
\end{center}
\end{table}

\begin{figure}[t]
\centering
\includegraphics[width=0.7\textwidth]{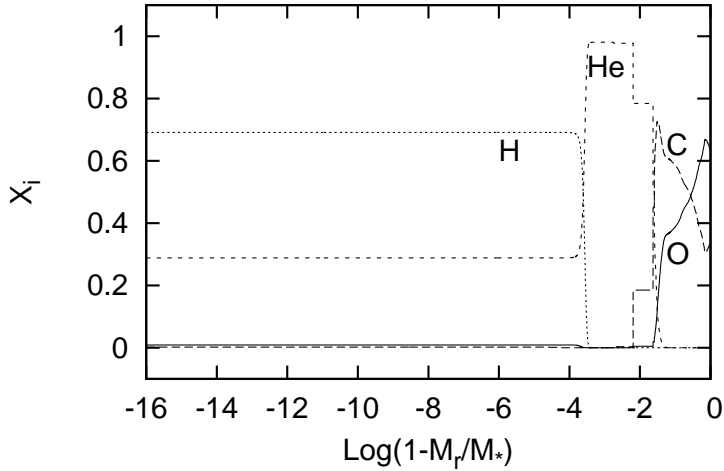}
\caption{Abundance of a 0.614 $M_{\odot}$ white dwarf model evolved from 2.8 $M_{\odot}$ main sequence star.}
\label{finger1}
\end{figure}

\begin{figure}[t]
\centering
\includegraphics[width=0.7\textwidth]{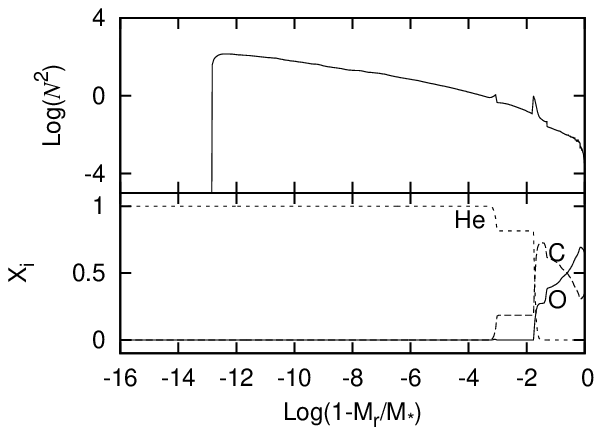}
\caption{Diagram of abundance and Brunt-V\"ais\"al\"a frequency for a DBV star. The model parameters are $T_{eff}$ = 25000 K, $M_{*}$ = 0.600 $M_{\odot}$, and log($M_{He}/M_{*}$) = -3.0.}
\label{finger2}
\end{figure}

\texttt{MESA} is a new program to do star evolutions, which can evolve stars from pre-main sequence to white dwarf stage. The cores of white dwarfs evolved by \texttt{MESA} are results of thermonuclear burning. The opacities are from Cassisi et al. (2007). The thermonuclear reaction rates are from Caughlan \& Fowler (1988) and Angulo et al. (1999), seeing Paxton et al. (2011) for more details. We download and install the 4298 version of \texttt{MESA}. A module named 'make\_co\_wd' (mesa/star/test\_suite/make\_co\_wd) is used to evolve stars from main sequence stage to white dwarf stage. With default input settings (mixing length parameter $\alpha$ = 2.0 and metal abundance Z = 0.02), we evolve stellar mass for main sequence stars from 2.0 $M_{\odot}$ to 5.0 $M_{\odot}$, as shown in the first column in Table 1. On the Hertzsprung-Russell diagram, we trace the luminosity. When those stars entering the white dwarf cooling track, we stop their evolutions. The corresponding white dwarf masses are showed in the second column in Table 1. Those white dwarf models evolved by 'make\_co\_wd' usually have a thick helium layer and hydrogen atmosphere. In Fig. 1, we show abundance of a 0.614 $M_{\odot}$ white dwarf model evolved from 2.8 $M_{\odot}$ main sequence. We take the C/O core ($M_{core}$ = 0.595 $M_{\odot}$) out. With this method, those core masses are listed in the third column in Table 1. In Fig. 1, we can see that there is a He/C envelope above the C/O core. We take the location of He abundance being 0.5 as the boundary of envelope/core. We define $M_{en}$ as the mass of He/C envelope. Log($M_{en}/M_{core}$) equals -1.754 for the model in Fig. 1. In the envelope, $X_{C(en)}$ is defined as carbon abundance and $X_{He(en)}$ is helium abundance. In Fig. 1, $X_{C(en)}$ equals 0.185 and $X_{He(en)}$ equals 0.785. The other 3\% are abundances of $^{16}O$, $^{18}O$, $^{20}Ne$, $^{22}Ne$, $^{24}Mg$, and $^{28}Si$. For the helium layer in Fig. 1, we can also see that the He abundance is a little smaller than 1.000. However, we only calculate $^{1}H$(H), $^{4}He$(He), $^{12}C$(C), and $^{16}O$(O) by \texttt{WDEC}. For approximate calculations, we take $X_{C(en)}$ = 0.185 and $X_{He(en)}$ = 0.815 for the model, as shown in the fifth column in Table 1.

\texttt{WDEC} is an old program to do white dwarf evolutions, which is convenient to evolve grids of white dwarf models. The cores of white dwarfs evolved by \texttt{WDEC} are artificially constructed previously, such as full C, full O, homogeneous C/O profiles and so on. Therefore, we try to insert the core profiles of white dwarfs involved by \texttt{MESA} into \texttt{WDEC} to evolve grids of white dwarf models. The equation of state are from Lamb (1974) and Saumon et al. (1995). The radiative opacities and conductive opacities are from Itoh et al. (1983). The mixing length theory is from B\"{o}hm \& Cassinelli (1971) and Tassoul, Fontaine \& Winget (1990). The ratio of mixing length to pressure scale height ($\alpha$), for DBV star, is usually adopted around 1.25 (Bergeron et al. 2011, Montgomery 2007, Koester 2010). We adopt $\alpha$ = 1.25 for calculations.

With \texttt{MESA} evolutions, we obtain a grid of white dwarf models. The core mass, corresponding envelope mass, carbon abundance, and approximate helium abundance are showed in Table 1. Each core is used to evolve a group of white dwarfs by \texttt{WDEC}. For example, the 0.595 $M_{\odot}$ core is used to evolve white dwarfs from 0.580 $M_{\odot}$ to 0.605 $M_{\odot}$. The mass range of white dwarf models evolved by \texttt{WDEC} is showed in the fourth column in Table 1.

After \texttt{MESA} evolutions, we take out the structure parameters, including mass, radius, luminosity, pressure, temperature, entropy, and carbon abundance. Those parameters are inserted into the prototype models in \texttt{WDEC}. The oxygen abundance approximately equals 1.000 minus the carbon abundance. The envelope masses and C,He abundances in the envelope are confined by the last column in Table 1. The envelope masses are different in Table 1. There is even no envelope for white dwarfs evolved from 3.8 $M_{\odot}$ or 5.0 $M_{\odot}$ main sequence star. Core compositions and envelope abundances of white dwarfs evolved by \texttt{MESA} are used as input physics for \texttt{WDEC}. \texttt{WDEC} evolve white dwarfs from above 100,000 K to the DBV instability strip. A scheme of element diffusion derived by Thoul, Bahcall \& Loed (1994) is added into \texttt{WDEC} by Su et al. (2014). We take 500,000 years as the evolution step.

Grids of white dwarf models are evolved by \texttt{WDEC}. The stellar mass $M_{*}$ is from 0.550 $M_{\odot}$ to 0.850 $M_{\odot}$ with steps of 0.005 $M_{\odot}$. The effective temperature is from 21000 K to 30000 K with steps of 200 K. The envelope mass fraction is showed in the last column in Table 1 for each group of white dwarfs. The helium layer mass fraction (log($M_{He}/M_{*}$)) is from -6.0 to -3.0 with steps of 0.5. In Fig. 2, we show a diagram of abundance and Brunt-V\"ais\"al\"a frequency for an evolved DBV star. The model parameters are $T_{eff}$ = 25000 K, $M_{*}$ = 0.600 $M_{\odot}$, and log($M_{He}/M_{*}$) = -3.0. Its core compositions and envelope abundances are evolved from the white dwarf model in Fig. 1. The composition gradients in the low panel cause 'spikes' in the up panel. Those 'spikes' may cause mode trapping effect (Winget, Van Horn, \& Hansen 1981; Brassard et al. 1992). In addition, there is an extremely thin convection zone on the surface of the star. With those grids of DBV star models, we numerically solve the full equations of linear and adiabatic oscillation. Each eigenmode will be scanned out. Those calculated modes will be used to fit the observed modes of CBS 114.

\section{Model fittings on CBS 114}

In this section, we analyze the previous observed modes for CBS 114 and then make model fittings on CBS 114. According to different frequency splitting values and distributions of kinetic energy for corresponding fitting modes, we try to study the differential rotation effect for CBS 114.

\subsection{Mode identifications for CBS 114}

\begin{table}
\begin{center}
\caption{Mode identifications by HMW2002 and Metcalfe et al. (2005). $Fre.$ denotes the pulsating frequency in $\mu$Hz. $Per.$ is the corresponding period in second (s). $Fre.$ $Spl.$ is the frequency splitting value.}
\begin{tabular}{lllllll}
\hline
 ID              &$Fre.$ ($Fre.$ in HMW)  &$Fre.$ $Spl.$          &$Per.$ ($Per.$ in HMW) &Amp (Amp2001)  (Amp1988)   &$l$     & $m$  \\
\hline
                 &($\mu$Hz)               &($\mu$Hz)              &(s)                    &(mmag)                     &        &      \\
\hline
 $f_{1}$         & 1189.9                 &                       & 840.38                &     3.1                   &   1or2 & 0?   \\
                 &                        &                       &                       &                           &        &      \\
 $f_{2}$         & 1252.4                 &                       & 798.46                &     4.3                   &   1or2 & 0?   \\
                 &                        &                       &                       &                           &        &      \\
 $f_{3}$         & 1383.0                 &                       & 723.06                &     3.0                   &   1or2 & 0?   \\
                 &                        &                       &                       &                           &        &      \\
 $f_{4}$         & 1509.0                 &                       & 662.67                &     7.5                   &   1    & $-$1 \\
                 &                        &                 9.8   &                       &                           &        &      \\
 $f_{4}$         & 1518.8 (1518.75$^{-}$) &                       & 658.44 (658.43)       &    11.8 (16-33)  (33)     &   1    & 0    \\
                 &                        &                11.9   &                       &                           &        &      \\
 $f_{4}$         & 1530.7                 &                       & 653.30                &     9.0                   &   1    & $+$1 \\
                 &                        &                       &                       &                           &        &      \\
 $f_{5}$         & 1601.2                 &                       & 624.55                &     4.4                   &   1    & $-$1 \\
                 &                        &                10.9   &                       &                           &        &      \\
 $f_{5}$         & 1612.1 (1613.12)       &                       & 620.31 (619.91)       &    15.4 (10-17)  (15)     &   1    & 0    \\
                 &                        &                10.6   &                       &                           &        &      \\
 $f_{5}$         & 1622.7                 &                       & 616.27                &    13.3                   &   1    & $+$1 \\
                 &                        &                       &                       &                           &        &      \\
 $f_{6}$         & 1729.8 (1729.73)       &                       & 578.09 (578.13)       &     7.0 (22-37)  (18)     &   1or2 & 0?   \\
                 &                        &                       &                       &                           &        &      \\
 $f_{7}$         & 1829.6 (1835.64$^{+-}$)&                17.64? & 546.56 (544.77)       &     3.0 ($<$4.2) (13)     &   1    & 0?   \\
                 &                        &                       &                       &                           &        &      \\
 $f_{8}$         & 1963.2 (1969.60)       &                 6.40  & 509.37 (507.71)       &     5.5 (11-17)  ($<$5)   &   1    & 0?   \\
                 &                        &                       &                       &                           &        &      \\
 $f_{9}$         & 2086.5                 &                       & 479.27                &     4.8                   &   1    & $-$1?\\
                 &                        &                 7.1   &                       &                           &        &      \\
 $f_{9}$         & 2093.6                 &                       & 477.65                &     4.9                   &   1    & 0?   \\
                 &                        &                       &                       &                           &        &      \\
$f_{10}$         & 2317.7 (2306.38$^{+}$) &                       & 431.45 (433.58)       &     8.9 ( 4-12)  (13)     &   1or2 & 0?   \\
                 &                        &                       &                       &                           &        &      \\
$f_{11}$         & 2510.0 (2509.89$^{-}$) &                       & 398.41 (398.42)       &     6.8 ( 4-10)  ( 9)     &   1    & 0?   \\
                 &                        &                 5.2   &                       &                           &        &      \\
$f_{11}$         & 2515.2                 &                       & 397.59                &     3.2                   &   1    & $+$1?\\
\hline
\end{tabular}
\end{center}
\end{table}

In Table 2, we show mode identifications for CBS 114 by HMW2002 and Metcalfe et al. (2005). HMW2002 identified 7 independent modes. They were recovered by mode identifications of Metcalfe et al. (2005). In the second column, $Fre.$ denotes the frequencies identified by Metcalfe et al. (2005). In the parentheses, $Fre.$ in HMW denotes the frequencies identified by HMW2002. The minus sign means that the negative daily alias ($f_{i}$ - 11.60 $\mu$Hz) may also be the correct eigenfrequency. The plus sign means that the positive daily alias ($f_{i}$ + 11.60 $\mu$Hz) may also be the correct eigenfrequency, as reported in Table 2 of HMW2002. Amp2001 is the amplitude of modes identified by HMW2002. Amp1998 is the amplitude of modes re-analyzed by HMW2002 based on the observations of Winget \& Claver (1988, 1989). Metcalfe et al. (2005) treated $f_{10}$ + 11.60 $\mu$Hz as the eigenmode. Beside those 7 modes, Metcalfe et al. (2005) also identified 4 new ones, $f_{1}$, $f_{2}$, $f_{3}$, and $f_{9}$.

Metcalfe et al. (2005) identified 2 triplets $f_{4}$, $f_{5}$ and 2 doublets $f_{9}$, $f_{11}$, as shown in Table 2. In addition, we also treat modes of $f_{7}$ and $f_{8}$ as doublets, which are identified by HMW2002 and Metcalfe et al. (2005) together. The frequency splitting value is showed in the third column in Table 2. The mode $f_{7}$ identified by HMW2002 may be three values, 1824.04 $\mu$Hz, 1835.64 $\mu$Hz, and 1847.24 $\mu$Hz. Therefore, the frequency splitting for $f_{7}$ may be 5.56 $\mu$Hz, 6.04 $\mu$Hz, and 17.64 $\mu$Hz. Those frequency splitting values in Table 2 are scattered, which is very important to study the differential rotation.

In the fifth column in Table 2, we can see that the amplitude of $f_{4}$, $f_{6}$, and $f_{8}$ are decreasing from 2001 to 2004. The phenomenon is not unique, such as the DBV star GD 358 (Kepler et al. 2003). It may be caused by convective driving mechanism (Dupret et al. 2008). In the sixth column, we show an $l$ identification. If triplets or doublets are observed, the modes will be identified as $l$ = 1. There are $l$ = 2 modes observed on the prototype DBV GD 358 (Kepler et al. 2003) and on the $KEPLER$ DBV KIC 8626021 (Bischoff-Kim et al. 2014). We assume the singlets as $l$ = 1 or 2 modes. The $l$ $\geq$ 3 modes suffer from significant geometric cancelation (Dziembowski 1997).

Since the modes identified by Metcalfe et al. (2005) recover the previous identified 7 independent modes, we use these modes to constrain fitting models. For two triplets, we can see that the $m$ = 0 modes have higher amplitudes. Therefore, we assume that the singlets are $m$ = 0 modes. The higher amplitude modes for doublets are $m$ = 0 modes. At last, there are 6 $l$ = 1, $m$ = 0 modes and 5 $l$ = 1 or 2, $m$ = 0 modes identified in Table 2. Totally 11 modes are used to constrain fitting models.

\subsection{Fitting results for CBS 114}

\begin{figure}[t]
\centering
\includegraphics[width=0.7\textwidth]{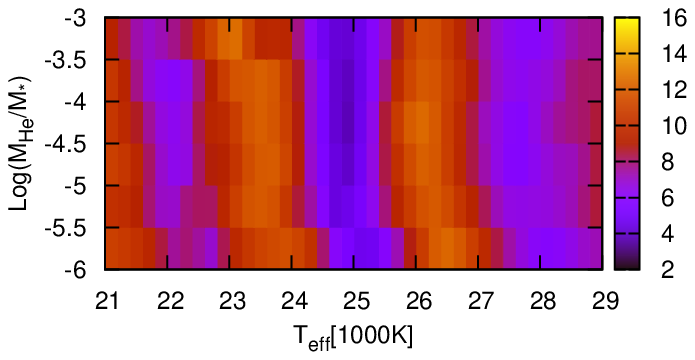}
\caption{Color residual diagram for models of $M_{*}$ = 0.740 $M_{\odot}$. The ordinate denotes the helium layer mass. The abscissa denotes the effective temperature. The color denotes the residual. The bluer the color, the smaller the residual.}
\label{finger3}
\end{figure}

\begin{figure}[t]
\centering
\includegraphics[width=0.7\textwidth]{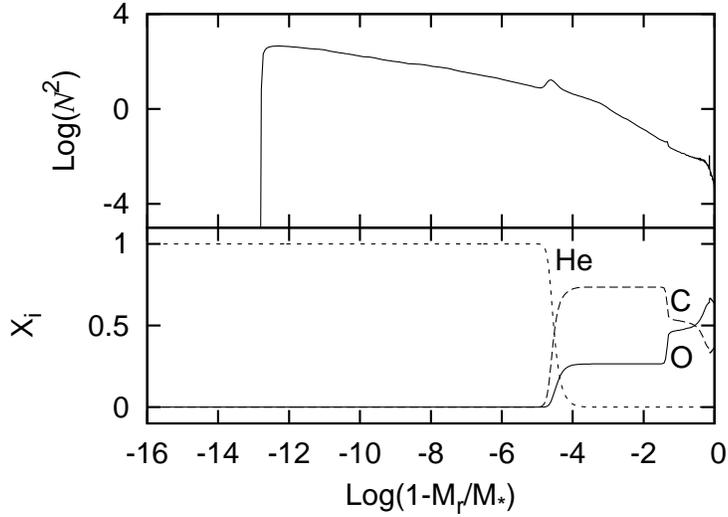}
\caption{Diagram of abundance and Brunt-V\"ais\"al\"a frequency for the best-fitting model. The model parameters are $T_{eff}$ = 25000 K, $M_{*}$ = 0.740 $M_{\odot}$, and log($M_{He}/M_{*}$) = -4.5.}
\label{finger4}
\end{figure}

\begin{table}
\begin{center}
\caption{The best-fitting models for asteroseismological study by HMW2002, Metcalfe et al. (2005), us, and spectroscopic study by Beauchamp et al. (1999), Kleinman et al. (2013).}
\begin{tabular}{lccccccccccc}
\hline
                       &$T_{eff}$(K)    &log$g$          &$M_{*}$/$M_{\odot}$  &log($M_{He}/M_{*}$)  \\
\hline
HMW2002(C-core)        &24600           &                &0.655                &-3.96                \\
HMW2002(O-core)        &25800           &                &0.640                &-3.96                \\
HMW2002(C/O core)      &21000           &                &0.730                &-6.66                \\
Metcalfe et al.(pure C)&25800           &                &0.630                &-5.96                \\
This paper             &25000           &8.287           &0.740                &-4.5                 \\
Beauchamp et al.(no H) &26200           &8.00            &                     &                     \\
Beauchamp et al.(H)    &23300           &7.98            &                     &                     \\
Kleinman et al.        &25754$\pm$368   &7.95$\pm$0.037  &                     &                     \\
\hline
\end{tabular}
\end{center}
\end{table}

We use the grids of DBV star models to fit the 11 identified $m$ = 0 modes. An usual residual formula is introduced by
\begin{equation}
\sigma_{RMS}=\sqrt{\frac{\sum_{1}^{n_{obs}}(Per._{\rm mod}-Per._{\rm obs})^{2}}{n_{obs}}}.
\end{equation}
\noindent In Eq.\,(3), $n_{obs}$ is the number of observed modes. In this paper, $n_{obs}$ is 11. $Per._{\rm mod}$ is the calculated periods and $Per._{\rm obs}$ is the observed periods. The model of smallest $\sigma_{RMS}$ is considered as the best-fitting one.

In Fig. 3, we show the color residual diagram for the fitting results. The bluer the color, the smaller the residual. The smallest $\sigma_{RMS}$ is 2.65\,s. A best-fitting model is selected. The model parameters are $T_{eff}$ = 25000 K, $M_{*}$ = 0.740 $M_{\odot}$, and log($M_{He}/M_{*}$) = -4.5. The model has gravitational acceleration of log$g$ = 8.287. In Table 1, we can see that the 0.740 $M_{\odot}$ white dwarf is from 3.8 $M_{\odot}$ main sequence star, which has no He/C envelope. In Fig. 4, we show a diagram of abundance and Brunt-V\"ais\"al\"a frequency for the best-fitting model. In the low panel, there is a C/O envelope above the C/O core, instead of a He/C envelope.

In Table 3, we show best-fitting models of asteroseismological study by HMW2002, Metcalfe et al. (2005), us and spectroscopic study by Beauchamp et al. (1999), Kleinman et al. (2013). Fitting 7 independent modes with C-core and O-core white dwarfs, HMW2002 obtained their similar best-fitting models. The parameters are $T_{eff}$ = 24600 K, $M_{*}$ = 0.655 $M_{\odot}$, log($M_{He}/M_{*}$) = -3.96 and $T_{eff}$ = 25800 K, $M_{*}$ = 0.640 $M_{\odot}$, log($M_{He}/M_{*}$) = -3.96 respectively. While, with C/O core white dwarfs, they obtained a best-fitting model of low effective temperature (21000 K), massive stellar mass (0.730 $M_{\odot}$), and thin helium atmosphere (log($M_{He}/M_{*}$) = -6.66). Fitting 11 independent modes with pure carbon core white dwarfs, Metcalfe et al. (2005) obtained a best-fitting model of $T_{eff}$ = 25800 K, $M_{*}$ = 0.630 $M_{\odot}$, and log($M_{He}/M_{*}$) = -5.96. For spectroscopic work, Beauchamp et al. (1999) obtained a best no hydrogen model of $T_{eff}$ = 26200 K, log$g$ = 8.00 and an undetectable hydrogen model with $T_{eff}$ = 23300 K, log$g$ = 7.98. In addition, from the spectroscopic catalog of Sloan Digital Sky Survey (SDSS) Data Release 7 (DR7), Kleinman et al. (2013) obtained a best-fitting model of $T_{eff}$ = 25754$\pm$368 K, log$g$ = 7.95$\pm$0.037.

The effective temperature for our best-fitting model is 25000 K. It is close to the previous asteroseismological results of 24600 K, 25800 K, and the previous spectroscopic result of 25754$\pm$368 K. The stellar mass for our best-fitting model is 0.740 $M_{\odot}$. It is obviously more massive than previous results of pure C-core or pure O-core white dwarf fittings. The stellar mass 0.740 $M_{\odot}$ is close to 0.730 $M_{\odot}$. Both of them are from C/O core white dwarf fittings to CBS 114. However, the gravitational acceleration (8.287) for out best-fitting model is obviously larger than the previous spectroscopic results, as shown in Table 3.

\subsection{Differential rotations for CBS 114}

\begin{table}
\begin{center}
\caption{Table of fitting results. $Per._{\rm obs}$ and $Per._{\rm mod}(l,k)$ denote observed periods and calculated periods respectively. $Kin_{core}$ and $Kin_{He}$ denote percentage of kinetic energy distributed in C/O core and He layer respectively.}
\begin{tabular}{lllllll}
\hline
ID              &$Per._{\rm obs}$($Fre.$ $Spl.$)&$Per._{\rm mod}(l,k)$      &$(Kin_{core},Kin_{He})$    &$Per._{\rm obs}$-$Per._{\rm mod}(l,k)$\\
\hline
                &(s)             ($\mu$Hz)      &(s)                        &                           &(s)                                   \\
\hline
$f_{11}$        &398.41          ( 5.2 )        &396.681  (1, 8)            &(44\%, 56\%)               &$\,$ 1.729                            \\
$f_{10}$        &431.45                         &432.710  (2,17)            &                           &$\,$-1.260                            \\
                &                               &434.502  (1, 9)            &(78\%, 22\%)               &                                      \\
$f_{9}$         &477.65          ( 7.1 )        &474.397  (1,10)            &(66\%, 34\%)               &$\,$ 3.253                            \\
$f_{8}$         &509.37          ( 6.40)        &505.517  (1,11)            &(52\%, 48\%)               &$\,$ 3.867                            \\
$f_{7}$         &546.56          (17.64?)       &544.374  (1,12)            &(74\%, 26\%)               &$\,$ 2.186                            \\
$f_{6}$         &578.09                         &583.567  (2,24)            &                           &$\,$-5.477                            \\
                &                               &588.187  (1,13)            &(48\%, 52\%)               &                                      \\
$f_{5}$         &620.31          (10.6,10.9)    &617.807  (1,14)            &(67\%, 33\%)               &$\,$ 2.503                            \\
$f_{4}$         &658.44          (11.9, 9.8)    &659.540  (1,15)            &(68\%, 32\%)               &$\,$-1.100                            \\
                &                               &692.760  (1,16)            &(62\%, 38\%)               &                                      \\
$f_{3}$         &723.06                         &724.714  (1,17)            &(76\%, 24\%)               &$\,$-1.654                            \\
                &                               &763.581  (1,18)            &(67\%, 33\%)               &                                      \\
$f_{2}$         &798.46                         &800.004  (1,19)            &(60\%, 40\%)               &$\,$-1.544                            \\
$f_{1}$         &840.38                         &840.123  (1,20)            &(73\%, 27\%)               &$\,$ 0.257                            \\
\hline
\end{tabular}
\end{center}
\end{table}

\begin{figure}[t]
\centering
\includegraphics[width=0.7\textwidth]{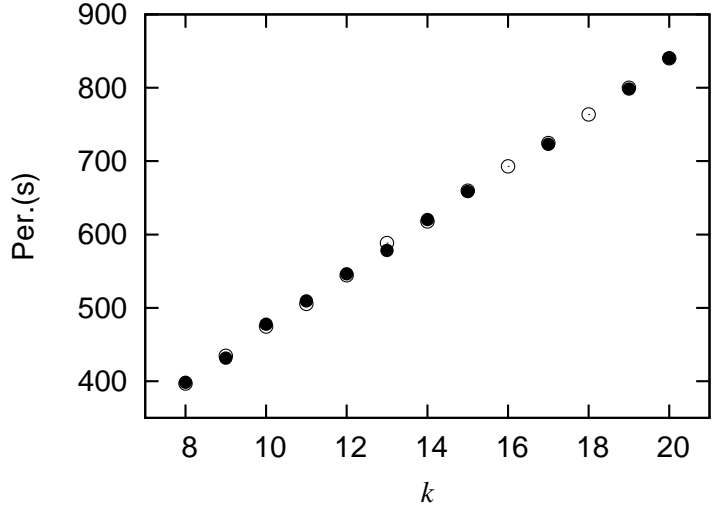}
\caption{Diagram of periods versus radial orders. The filled dots are the identified 11 eigenperiods. The open dots are the calculated $l$ = 1 modes for the best-fitting model. The radial order is from 8 to 20 for the calculated modes, which can be used as relative radial orders for the observed modes.}
\label{finger5}
\end{figure}

In Table 4, we show observed periods, calculated periods, fitting errors, and percentage of kinetic energy distributions respectively. The percentage of kinetic energy distributed in C/O core and He layer is calculated by the equation of
\begin{equation}
\frac{Kin_{core}}{Kin_{He}}=\frac{{4\pi\int_{0}^{R_{He/C}}[(|\tilde{\xi}_{r}(r)|^2+l(l+1)|\tilde{\xi}_{h}(r)|^2)]\rho_{0}r^2dr}}
                                 {{4\pi\int_{R_{He/C}}^{R}[(|\tilde{\xi}_{r}(r)|^2+l(l+1)|\tilde{\xi}_{h}(r)|^2)]\rho_{0}r^2dr}}.
\end{equation}
\noindent In Eq.\,(4), $\rho_{0}$ is the local density and $R_{He/C}$ is the location of He/C interface. For the best-fitting model, $R_{He/C}$ locates at log(1-$M_{r}/M_{*}$) = -4.5. $\tilde{\xi}_{r}(r)$ is the radial displacement and $\tilde{\xi}_{h}(r)$ is the horizontal displacement. The 11 observed modes are fitted by 9 $l$ = 1 modes and 2 $l$ = 2 modes. The mode $f_{10}$ is fitted by 432.710 s ($l$=2,$k$=17) in Table 4, which can also be fitted by the $l$ = 1 mode 434.502 s ($l$=1, $k$=9) with a small fitting error. However, the mode of $f_{6}$ can be only fitted by an $l$ = 2 mode. Even fitted by an $l$ = 2 mode, the fitting error is also not small (-5.477\,s). We doubt that whether $f_{6}$ is a $m$ = 0 component.

Assuming those 11 modes as $l$ = 1 and $m$ = 0, Metcalfe et al. (2005) obtained their best-fitting model with $\sigma_{RMS}$ = 2.33\,s. They did not show detailed calculated modes and fitting errors. We do not know their fitting result for $f_{6}$. For those 11 modes, there is a good mean period spacing of 36.534\,s. HMW2002 showed a mean period spacing of 37.1$\pm$0.7\,s. In Fig. 5, the filled dots show the observed 11 modes. They present a good straight line versus the calculated radial orders. The open dots show the calculated $l$ = 1 modes of the best-fitting model. On the whole, the open dots match the filled dots well. For details, $f_{6}$ is badly fitted. However, it is impossible to rule out an $l$ = 2 mode standing in the $l$ = 1 sequence. It is also impossible to rule out that $f_{6}$ is a $m$ $\neq$ 0 mode. Therefore, a relative reliable identification on $f_{6}$ requires more observations.

The percentage of kinetic energy distributions for $l$ = 1 modes are showed in the fourth column in Table 4. In the parenthesis of the second column, the frequency splitting values are showed again. We notice that the observed mode $f_{11}$ with 5.2 $\mu$Hz frequency splitting corresponds to the calculated $l$=1, $m$=8 mode. This mode has 56\% of kinetic energy distributed in He layer. The observed mode $f_{8}$ with 6.40 $\mu$Hz frequency splitting corresponds to the calculated $l$=1, $m$=11 mode. This mode has 48\% of kinetic energy distributed in He layer. While, $f_{4}$ with 11.9 $\mu$Hz, 9.8$\mu$Hz frequency splitting corresponds to $l$=1, $m$=15 mode. This mode has 32\% of kinetic energy distributed in He layer. It seems that the smaller the frequency splitting value for an observed mode, the more the kinetic energy distributed in He layer for a calculated mode. The phenomenon may be caused by modes partly trapped in He layer or C/O core. Rotation can cause frequency splitting effect, as shown in Eq.\,(2). Therefore, we suggest that the C/O core may rotate faster than the He layer for CBS 114. This is a preliminary inference. The frequency splitting values are scattered, being 5.2 $\mu$Hz, 6.40$\mu$Hz, 7.1$\mu$Hz, 9.8$\mu$Hz, 10.6$\mu$Hz, 10.9$\mu$Hz, 11.9$\mu$Hz. The frequency splitting values and the kinetic energy distributions show that the C/O core may rotate at least 2 ((11.9,9.8,10.6,10.9)/5.2 $\sim$ 2) times faster than the He layer. Kawaler, Sekii, \& Gough (1999) reported that the center rotated faster than the surface for PG1159 and the inner core might rotate rapidly for GD358. More observations are required in order to quantify the differential rotation. In addition, the mode $f_{7}$ is identified as 1829.6 $\mu$Hz by Metcalfe et al. (2005) and 1835.64$^{+-}$ by HMW2002. For the calculated mode, there are only 26\% of kinetic energy distributed in He layer. The frequency splitting value may be larger than 11.9 $\mu$Hz for $f_{7}$. Therefore, we suggest that 1835.64 + 11.60 $\mu$Hz may be the eigenfrequency and corresponding frequency splitting value may be 17.64 $\mu$Hz.

\section{Discussion and conclusions}

In this paper, we evolve grids of white dwarf models by \texttt{MESA} and then we insert those white dwarf cores into \texttt{WDEC}. This method has been used by us to evolve grids of DAV star models to study EC14012-1446 (Chen \& Li 2014). For DBV stars, we improve the method by extracting the He/C envelope from white dwarf models evolved by \texttt{MESA}. Metcalfe et al. (2005) made dual-site observations on CBS 114 and identified 11 independent modes. The $m$ = 0 modes in two triplets show higher amplitudes. Therefore, the higher amplitude modes in doublets are identified as $m$ = 0 modes. The singlets are identified as $m$ = 0 modes. Triplets and doublets are identified as $l$ = 1 and singlets are identified as $l$ = 1 or 2 by us.

Fitting the 11 identified modes, we obtain a best-fitting model of $T_{eff}$ = 25000 K, $M_{*}$ = 0.740 $M_{\odot}$, and log($M_{He}/M_{*}$) = -4.5. The residual is 2.65\,s and the gravitational acceleration is log$g$ = 8.287. The effective temperature is close to previous spectroscopic result of 25754$\pm$368 K. However, the gravitational acceleration is obviously larger than the spectroscopic result of 7.95$\pm$0.037. In fact, our evolved DBV star models are contradictory to the 'DB gap' phenomenon. In the future work, we will try to evolve DBV star models with a few hydrogen left in order to pass the 'DB gap'. The helium convective dilution effect will change those DA stars into DB stars when $T_{eff}$ cools down to 30000 K, as discussed in the introduction. Kleinman et al. (2004) reported that there were relative overabundance of DA stars inside the 'DB gap' according to the SDSS DR1. The DBV star models with few hydrogen atmosphere may have a chance to solve the problem of large gravitational acceleration.

At last, we study the frequency splitting values for observed modes and the kinetic energy distributions for best-fitting modes. The observed modes with large frequency splitting values correspond to the calculated modes with much kinetic energy distributed in C/O core. The frequency splitting values of $f_{4}$ and $f_{5}$ are basically 2 times of which of $f_{11}$, as shown in Table 4. We suggest that the C/O core may rotate at least 2 times faster than the He layer for CBS 114.

\section{Acknowledgment}

This work is supported by the National Natural Science Foundation of China (Grant No.11563001) and the Research Fund of Chuxiong Normal University (XJGG1501). We are very grateful to H. Shu and C. Y. Ding for their kindly discussion and suggestions.

\label{lastpage}
\end{document}